\documentclass[twocolumn,amsmath,amssymb,amsfonts,aps,prb,groupedaddress]{revtex4-1}
\usepackage{graphicx}% Include figure files
\usepackage{subfigure} % Make multi-figure figures
\usepackage{dcolumn}% Align table columns on decimal point
\usepackage{bm}% bold math
\usepackage{color}

\begin{document}

\title{Substrate influence on the plasmonic response of clusters of spherical nanoparticles}


\author{P. A. Letnes}
\email[]{paul.anton.letnes (at) gmail.com}
%\thanks{}
\author{I. Simonsen}
\affiliation{Department of Physics, The Norwegian University of Science and Technology (NTNU), NO-7491 Trondheim, Norway}
\author{D. L. Mills}
\affiliation{Department of Physics and Astronomy, University of California, Irvine, California 92697, USA}

\date{\today}

\begin{abstract}
The plasmonic response of nanoparticles is exploited in many subfields of science and engineering to enhance optical signals associated with probes of nanoscale and subnanoscale entities. We develop a numerical algorithm based on previous theoretical work that addresses the influence of a substrate on the plasmonic response of collections of nanoparticles of spherical shape. Our method is a real space approach within the quasi-static limit that can be applied to a wide range of structures. We illustrate the role of the substrate through numerical calculations that explore single nanospheres and nanosphere dimers fabricated from either a Drude model metal or from silver on dielectric substrates, and from dielectric spheres on silver substrates.
\end{abstract}

%\keywords{}

\maketitle

\section{\label{sec:introduction} Introduction}

Currently there is great interest in the use of the plasmonic response of tailored metallic substrates and other structures for the purpose of enhancing electric fields of laser beams in their near vicinity.  Enhancements with origin in the excitation of collective plasmon modes can increase the field intensity by many orders of magnitude in the near vicinity of diverse systems. This phenomenon was first explored in the context of surface enhanced Raman scattering (SERS), wherein it was found that the Raman cross section of pyridine adsorbed on electrochemically roughened Ag surfaces can be enhanced by approximately six orders of magnitude relative to that realized for pyridine in solution~\cite{RevModPhys.57.783}. The field has evolved to the point where the Raman spectrum of single molecules can be detected through use of plasmon enhanced Raman probes~\cite{ShumingNie02211997,doi:10.1021/jp034632u}. Plasmonic enhancements can be used not only in the context of Raman spectroscopy, but more generally to enhance the cross section of diverse nonlinear optical processes~\cite{PhysRevB.27.4553,ichimura2003:rt}.

In the theoretical literature, one finds numerous studies of the plasmonic response of isolated nanoparticles of diverse shape~\cite{PhysRevLett.91.253902,PhysRevB.72.155412} along with metallic arrays of nanoparticles~\cite{Zou:2006fk,PhysRevB.68.245420}. So far as we know, virtually all such discussions explore nanoparticles and their arrays in free space~\cite{PhysRevLett.101.197401,Gerardy:1980lh}. Treatments of the free space response are appropriate for clusters of nanoparticles in solution, but commonly one is interested in particles and particle arrays on substrates. Then an issue is the influence of the nanoparticle-substrate interaction on the plasmonic response of the nanoparticles that reside on it.  Papers addressing particle-substrate interactions include the work of Yamaguchi et al.~\cite{Yamaguchi1974173}, which discussed particles above substrates in the dipole approximation. Work done by Ruppin~\cite{Ruppin1983108} and by Noguez et al.~\cite{PhysRevB.61.10427,doi:10.1021/jp066539m} also deal with sphere-substrate interactions, but only for a single nanoparticle. Mayergoyz et al. have studied the plasmon eigenfrequencies of nanosphere dimers and also cylindrical structures on a substrate~\cite{PhysRevB.72.155412}. Moreover, a recent study on the plasmonic response of cubical nanoparticle dimers\cite{doi:10.1021/nn101484a} reports on the dimer-substrate interactions in the SERS context.

Since the early 1970s, Bedeaux and Vlieger et al. have conducted numerous theoretical and numerical studies on the effects of particle-substrate and particle-particle interactions~\cite{bedeaux_book}. These studies have been concentrated around spherical or spheroidal particles on top of a substrate, or \emph{truncated} particles of such shapes on a substrate (used to model a finite contact angle). In this work, the particle-substrate interactions were taken into account to high multipolar order, while the particle-particle interactions were only calculated to dipolar or quadrupolar order, since their main concern was systems of low or finite particle coverage~\cite{bedeaux_book,Simonsen:2003aa}.

More recently, numerical studies based on Bedeaux and Vlieger's work have been carried out by Simonsen, Lazzari and co-workers for the purpose of in-situ inversion of experimental optical spectra obtained from growing thin granular metal films~\cite{Simonsen:2003aa,PhysRevB.61.772,PhysRevB.65.235424,Lazzari2002124}.

In this paper, we present a description of the influence of a substrate on the plasmonic response of non-periodic nanosphere arrays; through use of the Bloch theorem one may address periodic systems as well. We employ the quasi-static description of the response of the system. This proves adequate for objects whose linear dimensions are small compared to the wavelength of light~\cite{Jackson:1962aa}. In contrast to previous work, we consistently take into account higher order interactions between the nanospheres.

After we describe the formalism, we turn our attention to calculations that explore the influence of the substrate on the response of nanospheres and nanosphere dimers. Of interest is the discussion of ``hot-spot''-regions where at selected excitation frequencies one realizes very large field enhancements by virtue of the excitation of collective plasmon modes. For the case of two spheres in free space that are nearly in contact, one realizes a hot-spot at the point of closest contact between the spheres~\cite{Chu:2007ph,Chu:2008gd}. In this paper, for a nanosphere dimer near a dielectric substrate we find  ``moving hot-spots''. A small change in excitation frequency can cause the hot-spot to move from the point of nearest contact between the spheres, to the ``south poles'' of the spheres---the points on the spheres closest to the substrate. In recent work, two of the authors have discussed moving hot-spots in nanosphere clusters~\cite{Letnes:2010uq}.

The present study illustrates the role of the substrate in creating new hot-spots. We find  that if a dielectric sphere is in close proximity to a ``plasmonic active'' metallic substrate, the region around the south pole of the dielectric sphere becomes a hot-spot. A spatially localized ``potential well'' that can trap substrate plasmons is formed just under the dielectric sphere. Also, if a metallic sphere is placed close to a dielectric substrate, we find a collective plasmon localized near the south pole of the sphere. Thus, the interaction of nanospheres and structured arrays of such objects placed on substrates creates new hot-spots that can be exploited in diverse non-linear optical spectral probes of nanoscale and subnanoscale matter.

In this paper, Sec.~\ref{sec:Theory} is devoted to setting up a formalism that may be applied to any non-periodic structure of spherical nanoparticles that are located on, or near, a substrate, and Sec.~\ref{sec:Results-and-discussion} presents the results of our numerical studies of isolated nanospheres and nanosphere dimers on substrates. Section~\ref{sec:Concluding-remarks} contains concluding remarks.

\section{Theory} % (fold)
\label{sec:Theory}

\begin{figure}[bthp]
    \includegraphics{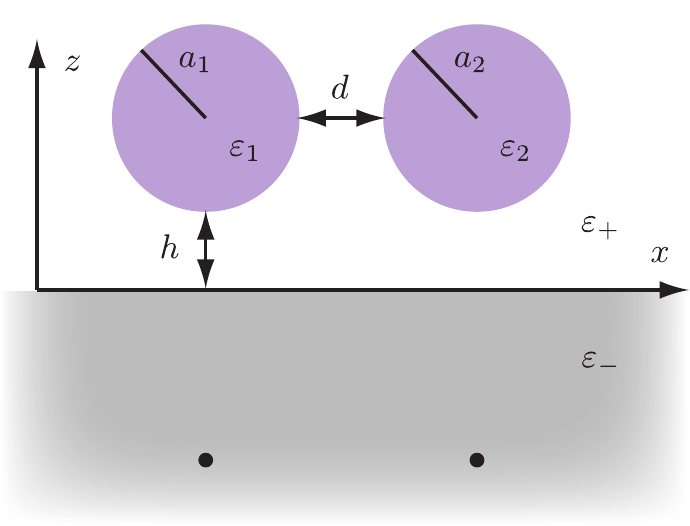}
    \caption{\label{fig:geometry}
    (Color online)
    An illustration of the system we consider in this paper, for the case where we have a nanosphere dimer. The substrate occupies the half space $z < 0$, $h$ is the distance between the ``south poles'' of the spheres and the substrate, and $d$ is their surface-surface separation. Sphere $j$ of the dimer has dielectric function $\varepsilon_j(\omega)$ and radius $a_j$. The medium above the substrate has dielectric function $\varepsilon_+(\omega)$ while that of the substrate is $\varepsilon_-(\omega)$. The two black dots represent schematically image multipoles in the substrate seen by an observer in the half space $z > 0$.}
\end{figure}

Even if the numerical calculations to be performed in this paper will focus on one or two nanoparticles, we will however present a more general formalism valid for a cluster of $N$ nanoparticles. For the case of the dimer, the geometry is illustrated in Fig.~\ref{fig:geometry}. The substrate is located in the half space $z < 0$ and it is characterized by the dielectric function $\varepsilon_-(\omega)$. The region above the substrate, $z > 0$, is assumed to be a non-absorbing dielectric characterized by the dielectric function $\varepsilon_+(\omega)$.

We consider a system consisting of $N$ non-overlapping nanospheres, located at arbitrariy positions. For each such sphere we embed a coordinate system $\mathcal{S}_j$, $j = 1,2,\ldots N$, so that the origin of $\mathcal{S}_j$ is located at the center of sphere $j$. With each coordinate system $\mathcal{S}_j$, we associate a position vector $\bm{r}_j = (r_j, \theta_j, \phi_j)$.

Our interest is in nanosphere arrays whose extent is small compared to the wavelength of light, so the electrostatic approximation suffices to describe the electric fields in its vicinity. Under this assumption, the Maxwell's equations are equivalent to the Laplace equation. Thus our task is to solve Laplace's equation for the electrostatic potential $\psi$,
\begin{align*}
    \bm{\nabla}^2 \psi = 0,
\end{align*}
subject to the appropriate boundary conditions on the surface of each sphere, and at the interface between the substrate and the rest of the system. As usual, the electric field is given by~\cite{Jackson:1962aa} $\bm{E} = -\bm{\nabla} \psi$.

We will assume that a spatially uniform electric field $\bm{E}_0$ of angular frequency $\omega$ is applied to the system and we analyze its response to this field. In what follows all dielectric functions that enter the analysis are the complex dielectric functions appropriate for the frequency $\omega$, though we suppress any explicit reference to $\omega$ in what follows. Hence, the electrostatic potential in the half space $z > 0$ can be written as
\begin{align}
\label{eq:plus-expansion}
        \psi_+ (\bm{r}) = -\bm{r} \cdot \bm{E}_0
            + \sum_{j = 1}^{N} \psi_j(\bm{r}_j)
            + \sum_{\bar{j} = 1}^{N} \psi_{\bar{j}} (\bm{r}_{\bar{j}}),
\end{align}
where $\psi_j$ is the electrostatic potential produced by the polarization charges in sphere $j$ and $\psi_{\bar{j}}$ the potential produced by its image, located in the half space $z < 0$. In the substrate ($z < 0$), the electrostatic potential takes the form
\begin{align}
\label{eq:minus-expansion}
        \psi_- (\bm{r}) = -\bm{r} \cdot \bm{E}_0^{T}
            + \sum_{j = 1}^{N} \psi_j^{T} (\bm{r}_j),
\end{align}
where $\psi_j^T$ is the electrostatic potential of sphere $j$ as seen by an observer in the region $z < 0$, and $\bm{E}_0^T$ is the applied field in the substrate. The various single sphere potential functions that enter Eqs.~\eqref{eq:plus-expansion} and~\eqref{eq:minus-expansion} may be expanded in the spherical harmonics. Using the shorthand notation $\sum_{lm} = \sum_{l=0}^\infty \sum_{m=-l}^l$ we have
\begin{subequations}
\label{eq:psi_j}
\begin{align}
    % Scattered...
    \psi_j (\bm{r}_j)
        &=
        \begin{cases}
            \sum_{lm} A^{(j)}_{lm} r_j^{-l-1}
                Y_l^m (\theta_j, \phi_j), &
                r_j \geq a_j, \\
            \sum_{lm}
                B^{(j)}_{lm} r_j^l
                Y_l^m (\theta_j, \phi_j), &
                r_j < a_j, \\
        \end{cases} \\
    % reflected...
    \psi_{\bar{j}} (\bm{r}_{\bar{j}}) &=
        \sum_{lm} A^{(j, R)}_{lm}
            r_{\bar{j}}^{-l-1}
            Y_l^m (\theta_{\bar{j}}, \phi_{\bar{j}}),
    % and transmitted potential.
    \end{align}
    and
    \begin{align}
     \psi_j^T (\bm{r}_j) &=
        \sum_{lm} A^{(j, T)}_{lm}
            r_j^{-l-1} Y_l^m (\theta_j, \phi_j), &
\end{align}
\end{subequations}
where the various $A_{lm}$ and $B_{lm}$ are expansion coefficients to be determined, and $a_j$ refers to the radius of sphere $j$. The symbol $Y_l^m$ refers to the spherical harmonic functions as described by Ref.~\onlinecite{Jackson:1962aa}. As discussed in Refs.~\onlinecite{bedeaux_book,PhysRevB.61.772}, the coefficents $A_{lm}^{(j,R)}$ and $A_{lm}^{(j,T)}$ are related to $A_{lm}^{(j)}$ through the boundary conditions at the interface $z = 0$. Simple image arguments supply the relation between these quantities. In particular, one finds that~\cite{bedeaux_book,PhysRevB.61.772}
\begin{subequations}
\label{eq:image-bc}
    \begin{align}
        A_{lm}^{(j, R)} &= (-1)^{l+m}
            \frac{\varepsilon_+ - \varepsilon_-}
                {\varepsilon_+ + \varepsilon_-}
            A_{lm}^{(j)},
    \end{align}
and
    \begin{align}
        A_{lm}^{(j, T)} &=  
            \frac{2 \varepsilon_+}
                {\varepsilon_+ + \varepsilon_-}
            A_{lm}^{(j)}.
    \end{align}
\end{subequations}
Equation~\eqref{eq:image-bc} insures that the boundary conditions on the substrate ($z = 0$) are automatically satisfied for any $A_{lm}^{(j)}$. Thus in what follows, we seek to solve for the coefficients $A_{lm}^{(j)}$ and $B_{lm}^{(j)}$ using the equations that follow from the boundary conditions at the surface of each nanosphere, i.e. where $r_j = a_j$.  
Through rearrangement of the equations following from the boundary conditions on the sphere surfaces, one can eliminate the coefficients $B_{lm}^{(j)}$. In Appendix~\ref{sec:Construction the equation system} the linear set of equations determining $A_{lm}^{(j)}$ and $B_{lm}^{(j)}$ are derived [cf. Eq.~\eqref{eq:eq-system}].

In Sec.~\ref{sec:Results-and-discussion}, we present a series of numerical studies of plasmon resonance phenomena for nanosphere monomers and dimers placed on a substrate. To this end, we must solve Eq.~\eqref{eq:eq-system}. In order to do so, we truncate the summations in Eq.~\eqref{eq:psi_j} and also the equation system in Eq.~\eqref{eq:eq-system} at $l = L$. The number of unknown coefficients in Eq.~\eqref{eq:eq-system} is then $N (L + 1)^2 - 1$. We use the same truncation limit for both the nanosphere-nanosphere interactions, as well as for the nanosphere-substrate interactions. The nanosphere-substrate interactions include both the interaction of a given nanosphere with its own image, but also the images of the other nanospheres. Note that this is the first time, to the best of our knowledge, that all particle-substrate \emph{and} particle-particle interactions have been taken consistently into account (to a given order). In several previous studies the interaction with the substrate has been taken into account to a high order, while the particle-particle interactions have been accounted for to dipolar or quadrupolar order~\cite{Simonsen:2003aa,Lazzari2002124,PhysRevB.65.235424,PhysRevB.61.772,bedeaux_book}. We shall see in the next section that the use of the dipole approximation (retention of only the terms with $l = 1$) in the particle-substrate interaction is very inaccurate from a quantitative point of view save for the case when the nanospheres are quite far from the substrate.

In passing we note that the formalism presented in this paper can be applied to extend the formalism used in Ref.~\onlinecite{PhysRevB.68.245420} to incorporate interactions of periodic structures with a substrate. One then has plasmon normal modes characterized by a wave vector $\bm{k}_{\parallel}$ parallel to the surface; one encounters only $(L+1)^2$ coefficients in this case, because the expansion coefficients of different nanoparticles are linked by the Bloch theorem. The quasi-static limit developed in this paper can be applied to the description of collective excitations whose wave vector is large compared to $\omega / c$, with $\omega$ the angular frequency of an excitation of interest, and $c$ the velocity of light in vacuum.

% section Theory (end)

\section{Results and discussion} % (fold)
\label{sec:Results-and-discussion}

In this section, we present a series of studies of the influence of a dielectric substrate on the plasmonic response of isolated nanospheres and nanosphere dimers. In addition, we find ``hot-spots'' created by plasmonic resonances between dielectric spheres and metallic substrates, as noted above. We shall also see that termination of the hierarchy of equations at the dipole ($L = 1$, see Ref.~\onlinecite{Yamaguchi1974173})  or quadrupole ($L = 2$) order provides a very poor quantitative description of interactions between particles, and  between the particles and the substrate. We remark that it is evident from earlier studies, which utilize a different methodology~\cite{Chu:2008gd}, that higher order harmonics must be included in the description of particle-particle interaction, since the fields associated with ``hot-spots'' are highly localized around the points of nearest contact. Thus one must retain spherical harmonics to high order to describe these features.

For the purpose of studying particle-substrate interactions, we first consider nanoparticles modeled by a dielectric function of the Drude form~\cite{Jackson:1962aa},
\begin{align}
    \label{eq:drudemodel}
    \varepsilon(\omega) = 1 -
        \frac{\omega_P^2}
            {\omega (\omega + \mathrm{i} \gamma)},
\end{align}
where $\omega_P$ is the plasma frequency and $\gamma$ is the inverse of the free carrier relaxation time. For the ambient material we have chosen vacuum, i.e. $\varepsilon_+ = 1$, and a dielectric substrate of $\varepsilon_- > 0$.  The virtue of model studies based on the form of Eq.~\eqref{eq:drudemodel} is that we may choose the relaxation rate $\gamma$ sufficiently small so that much detail is evident in the calculated results. For the Drude model parameters we assume $\omega_P = 3$ eV and $\gamma = 0.03$ eV. After our discussion of nanospheres consisting of Drude metal, we present results for geometries incorporating silver (Ag) nanoparticles. Among metals which exhibit plasmonic response in the visible part of the optical spectrum, the damping rate in Ag is modest and numerous experiments employ Ag based structures~\cite{Tamaru:2002cs}.
It should be remarked that the optical response of aluminium (Al) is described very well by the Drude model. Unfortunately the plasma frequency is very high, close to $15$ eV, so the interesting plasmonic resonances in Al based materials lie well into the ultraviolet. In our view, it would be of great interest to see experimental probes of structures that incorporate Al nanoparticles, with attention to the appropriate spectral range.

One possible indicator of ``plasmonic activity'' is the total dipole moment of one of our spherical objects. With $\bm{p}(\omega)$ being the dipole moment of a nanosphere at angular frequency $\omega$, we define the dimensionless dipole moment as
%, then what we shall plot is the dimensionless measure of dipole moment given by 
\begin{align*}
    \bm{\bar{p}} = \frac{\bm{p}}{a^3 \varepsilon_0 E_0},
\end{align*}
where $a$ is the radius of the sphere in question, and $\varepsilon_0$ is the vacuum permeability. In terms of our expansion coefficients, $A_{lm}$, the three Cartesian components of the dimensionless dipole moment are given by
\begin{subequations}
    \begin{align*}
        \bar{p}_x &=
            \sqrt{\frac{3}{8 \pi}} \frac{A_{1,-1} - A_{1,1}}{a^2}, \\
        \bar{p}_y &= -\mathrm{i}
             \sqrt{\frac{3}{8 \pi}} \frac{A_{1,-1} + A_{1,1}}{a^2}, \\
     \end{align*}
 and
     \begin{align*}
        \bar{p}_z &= \sqrt{\frac{3}{4\pi}} \frac{A_{10}}{a^2}.
    \end{align*}
\end{subequations}
Since the dipole moment in general is a complex vector quantity, the quantity we display in the figures below is the modulus of the total dipole moment given by
\begin{align*}
    \bar{p} (\omega)
    \equiv |\bm{\bar{p}}(\omega)|
    = \sqrt{\bm{\bar{p}}^\dagger \bm{\bar{p}}},
\end{align*}
where $\dagger$ symbolizes the Hermitian transpose. In our studies of the interaction of a single sphere with the substrate, we shall display the total dimensionless dipole moment, along with field enhancement factors for applied fields perpendicular to the substrate ($\bm{E}_{0}\parallel \bm{\hat{z}}$) as well as parallel to the substrate ($\bm{E}_{0}\parallel \bm{\hat{x}}$). Moreover, for the dimer illustrated in Fig.~\ref{fig:geometry} we shall present results for all three Cartesian components of the applied field.

\subsection{The Drude Monomer and Dimer} % (fold)
\label{sub:drude-monomer-dimer}

\begin{figure}[btp]
    \centering
    \includegraphics{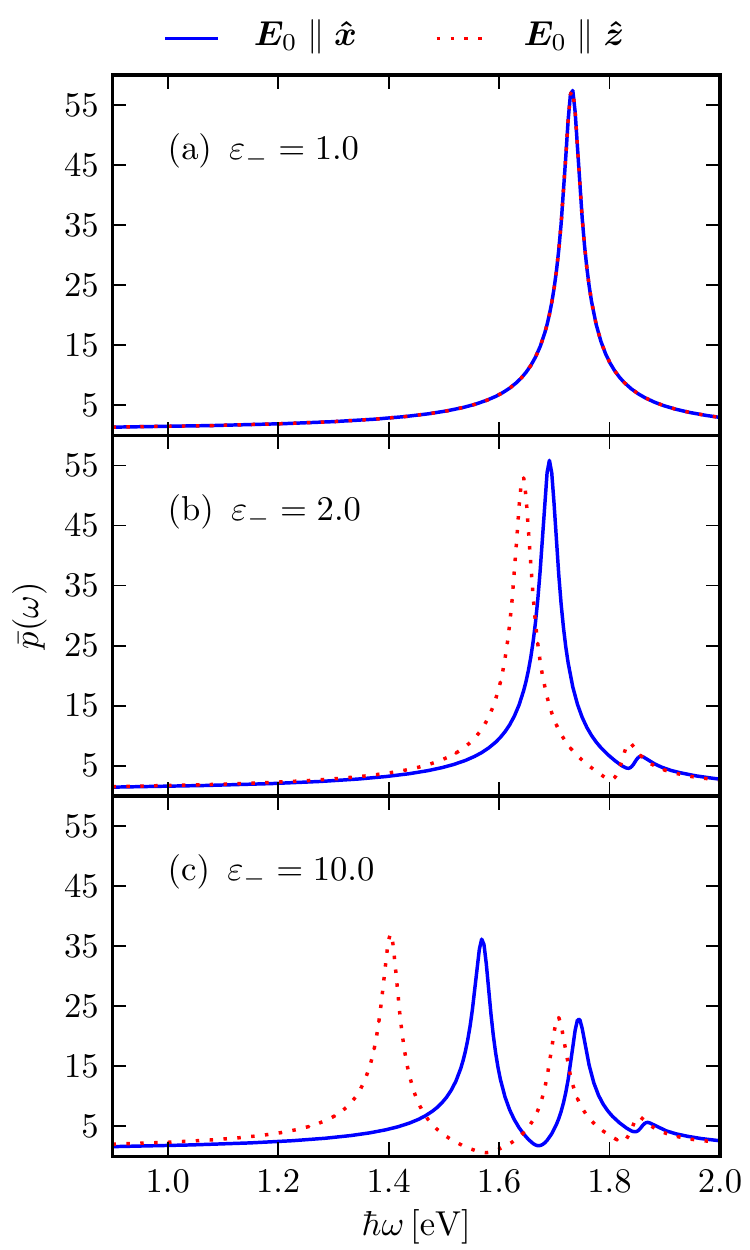}
    \caption{\label{fig:Drude-monomer-dipole}
    (Color online)
    The dimensionless dipole moment, $\bar{p}(\omega)$, for a Drude metal particle on a substrate of dielectric function (a) $\varepsilon_-=1$, (b) $\varepsilon_-=2$, and (c) $\varepsilon_-=10$. For the vacuum case, i.e. $\varepsilon_- = 1$, we obtain the Mie result at $\hbar\omega = \hbar\omega_P / \sqrt{3} \approx 1.73$ eV. For all plots, we have $h = 0.05 a, \, \omega_P = 3$ eV, $\gamma=0.03$ eV, and $L = 50$.}
\end{figure}

We begin by considering a single Drude sphere in vacuum ($\varepsilon_+ = 1$) located a distance $h = 0.05a$ above a substrate. In Fig.~\ref{fig:Drude-monomer-dipole}, we present numerical calculations of the dimensionless dipole moment  for three choices of the dielectric function of the substrate, $\varepsilon_- = 1, 2$, and $10$. The choice $\varepsilon_- = 1$ corresponds physically to the case where no substrate is present. We present results for two choices of the applied field: \emph{(i)} perpendicular to the substrate ($z$ direction, red dotted curves) and \emph{(ii)} parallel to the substrate ($x$ direction, blue solid curves). Since the system is invariant with respect to rotation about the $z$ axis, the response to an applied field parallel to the $y$ axis is identical to that shown for $x$ polarization.

\begin{figure}[tbhp]
    \centering
        \includegraphics{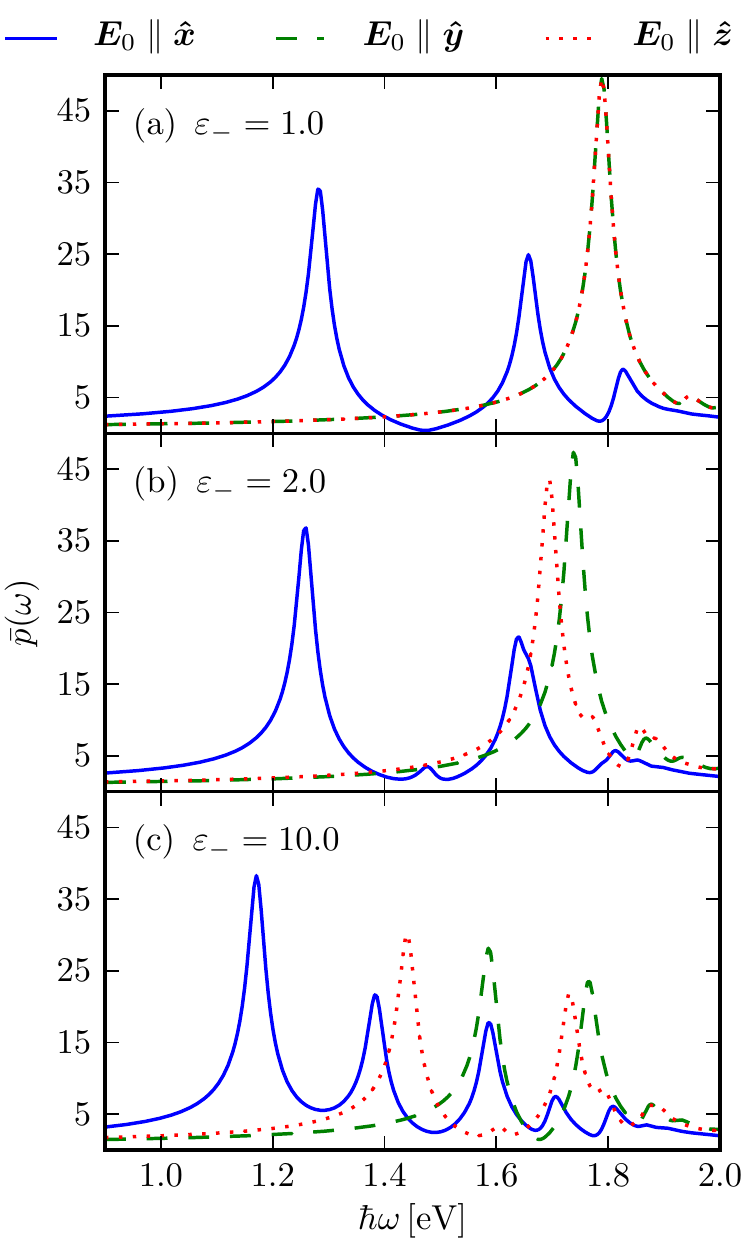}
    \caption{\label{fig:Drude-dimer-dipole}
    (Color online)
    The dimensionless dipole moment, $\bar{p}(\omega)$, for one of the particles in a Drude metal dimer, on top of a substrate of dielectric function (a) $\varepsilon_-=1$, (b) $\varepsilon_-=2$, and (c) $\varepsilon_-=10$. For all plots, we have $h = 0.05 a, \, d = 0.1 a, \, \omega_P = 3$ eV, $\gamma=0.03$ eV, and $L = 50$.}
\end{figure}

In Fig.~\ref{fig:Drude-monomer-dipole}(a), we present the response of the sphere in free space, from which the Mie resonance at the frequency $\hbar \omega_P / \sqrt{3} \approx 1.73$ eV is readily observed. While an isolated Drude metal sphere has a spectrum of multipole modes at angular frequencies $\omega_l = \omega_P \left( l / (2l + 1) \right)^{1/2}$ with $l = 1,2,3,\ldots$, only the dipole mode with $l = 1$ is excited by an applied field whose wavelength is large compared to the radius of the sphere. In the presence of a substrate, higher order modes may be excited by a spatially uniform applied field, as we shall see. These will appear at higher frequencies than the Mie resonance, as suggested by the fact that $\omega_{l + 1} > \omega_1$ for $l \geq 1$.

\begin{figure}[tbhp]
    \centering
        \includegraphics{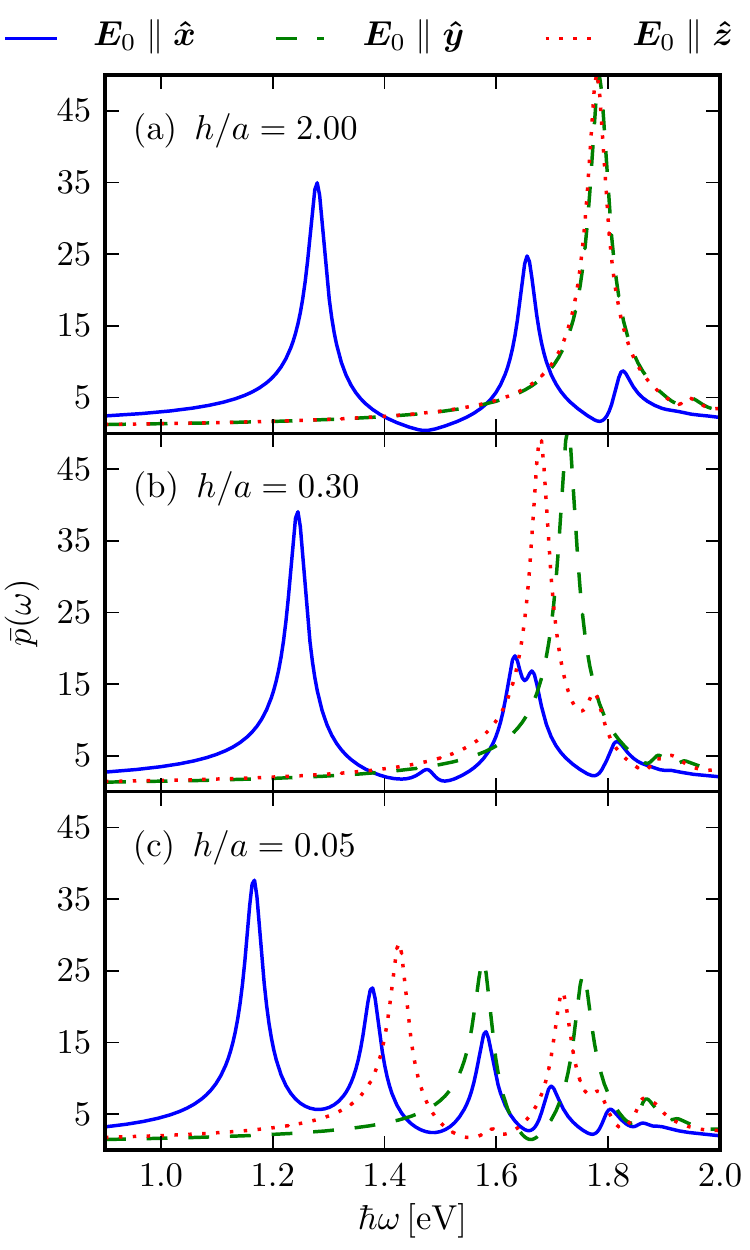}
    \caption{\label{fig:Drude-dimer-dipole-h-scan}
    (Color online)
    The plots show $\bar{p}(\omega)$ for one of the particles in a Drude metal dimer for different values of $h$. For all plots, we have $\varepsilon_- = 10$, $d = 0.1 a$, $\omega_P = 3$ eV, $\gamma=0.03$ eV and $L = 30$.}
\end{figure}

When the dielectric function of the substrate is $\varepsilon_- > \varepsilon_+$, the spectral response of the nanosphere is altered significantly. First for the case of modest dielectric function, $\varepsilon_- = 2$, the (former) Mie resonance remains the dominant spectral feature and it is red shifted by the proximity of the dielectric substrate to the nanosphere [Fig.~\ref{fig:Drude-monomer-dipole}(b)]. A substantial splitting of the modes is observed when the response to a field parallel to the substrate is compared to the response to a field perpendicular to it. We also see activation of a higher frequency mode.
One might be tempted to associate this with excitation of the quadrupolar mode with $l = 2$, but when the sphere is so close to the substrate, classification of the mode by the angular momentum quantum number is no longer accurate since a large number of $l$ modes are mixed together. We require $L$, the cutoff used in the hierarchy of equations displayed in Eq.~\eqref{eq:eq-system}, to be on the order of $30$ (or more) to obtain converged results.

If the substrate has a large dielectric function ($\varepsilon_- = 10$), then the response of the sphere is modified dramatically relative to the free-space case [Fig.~\ref{fig:Drude-monomer-dipole}(c)].
The splitting of the low frequency resonances, for parallel and perpendicular excitation, is now very large. For both orientations of the applied field, the oscillator strength of the next highest mode is comparable to the low frequency (dipole) mode. We also see a third mode in the spectrum, so the symmetry breaking provided by the substrate now asserts itself prominently in the response of the sphere.

The appearance of these higher order modes can be intuitively understood as follows. When we apply an electric field $\bm{E}_0$ to a nanosphere, it will generate local evanescent fields. When $h$ is small, some of the evanescent fields are reflected from the substrate, resulting in a non-uniform field around the sphere. This causes the simultaneous excitation of many different $l$ modes, meaning that the notion of discussing modes in multipolar terminology breaks down badly. The crosstalk between different $l$ modes is also the reason why we see higher order modes (e.g. ``quadrupole'' modes) in the dipole moment ($\bar{p}(\omega)$) of the spheres.

We now turn our attention to a discussion of the response of a Drude dimer, as shown in Fig.~\ref{fig:geometry}. The radiuses of the two spheres are both assumed to be equal to $a$. The distance between the spheres is $d = 0.1a$, and they are both placed a distance $h = 0.05a$ above the substrate. In Fig.~\ref{fig:Drude-dimer-dipole}, we depict the dimensionless dipole moment of one of the spheres in a dimer whose axis is parallel to the $x$ axis and, hence, to the substrate.

When the dimer is placed in free space [Fig.~\ref{fig:Drude-dimer-dipole}(a)] and the applied field is perpendicular to the dimer axis, there is one dominant resonance. This is the Mie resonance of the single sphere, slightly blueshifted due to the particle-particle interactions. In addition, a second weak mode shows up at higher frequency. In contrast, when the dimer is excited by a field parallel to the dimer axis (blue solid curve), we see a sequence of collective modes redshifted by large amounts from the isolated sphere Mie resonance. These results are in agreement with previous work on nanoparticle dimers in free space~\cite{Chu:2008gd}. As for the case of the single sphere on a substrate, the fields generated from one sphere causes higher order modes to be excited in the other sphere, and vice versa.

For a substrate with modest dielectric function [$\varepsilon_- = 2$, Fig.~\ref{fig:Drude-dimer-dipole}(b)] we see a splitting between the dominant collective modes excited by a field parallel to $\bm{\hat{z}}$ (red dotted curve) and that excited by a field parallel to $\bm{\hat{y}}$ (green dashed curve). This is to be expected since the presence of the substrate will break the rotational symmetry around the $x$ axis. If the dielectric function of the substrate is substantial [$\varepsilon_- = 10$, Fig.~\ref{fig:Drude-dimer-dipole}(c)] we observe dramatic differences between the spectral response for the three directions of the applied field. The shift of the lowest frequency mode for $\bm{E}_{0}\parallel \bm{\hat{x}}$ from the Mie resonance of the isolated sphere is particularly dramatic.

These results demonstrate that placing metallic nanosphere dimers over a substrate with large dielectric function will give rise to substantial field enhancements. Also, strong dipole moment enhancements can be achieved over a spectral range very large compared to that realized for a single isolated nanosphere. Thus, as this example illustrates, the interaction between structured nanoparticle arrays and a substrate of substantial dielectric function can allow one to design objects with a broad plasmonic spectral response. 

\begin{figure}[tbhp]
    \centering
        \includegraphics{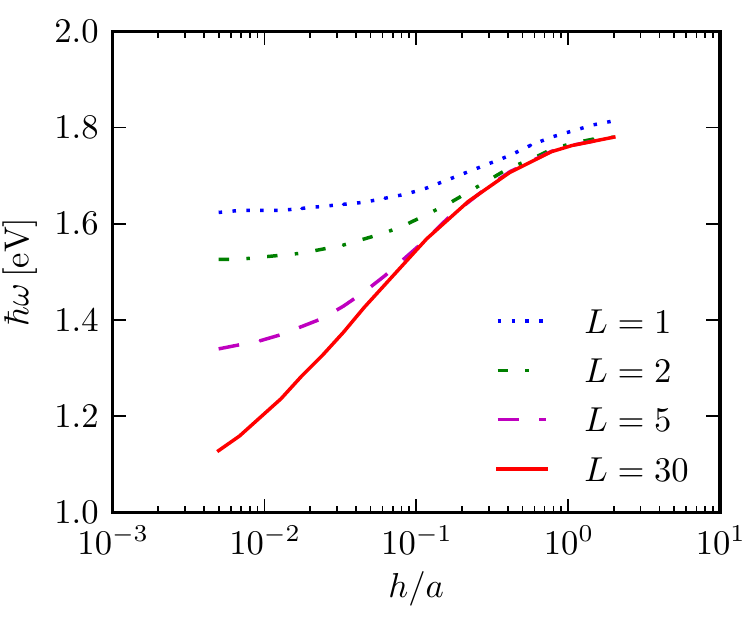}
    \caption{\label{fig:Drude-dimer-h-scan-Ez}
    (Color online)
    Position of the lowest energy resonance as a function of $h$, in the case where $\bm{E}_{0}\parallel\bm{\hat{z}}$. The blue dotted curve shows the result when the calculation is done in the dipole approximation (i.e. $L=1$), while the red solid curve $L=30$ show the converged results. For all cases, a Drude dimer with $d = 0.1 a$ was assumed, and the substrate dielectric function was $\varepsilon_- = 10$.}
\end{figure}

\begin{figure}[bthp]
    \centering
    \includegraphics{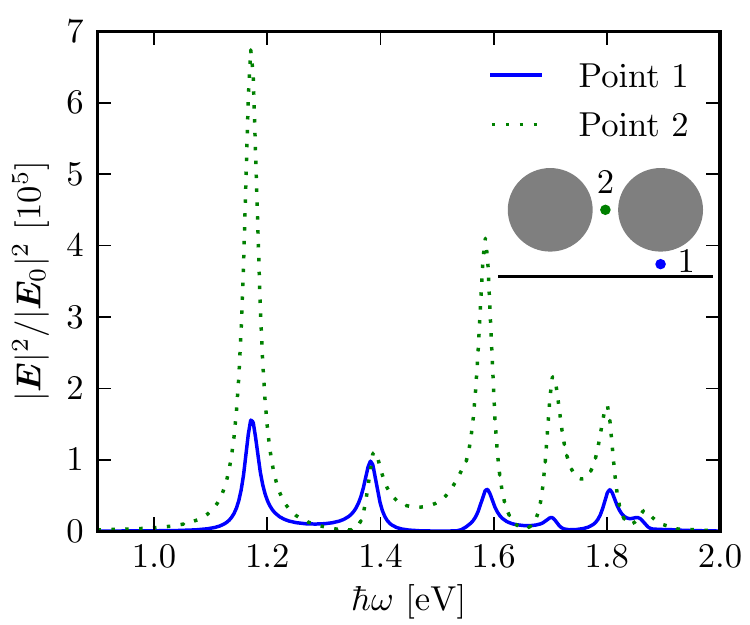}
    \caption{\label{fig:field-enhancement-Drude-gap}
    (Color online)
    The square of the electric field, $|\bm{E}|^2 / |\bm{E}_0|^2$, at point 1 (blue solid curve) and point 2 (green dotted curve) as a function of frequency, for the Drude dimer. The results are for the case where $\bm{E}_{0}\parallel \bm{\hat{x}}$, and the substrate dielectric function is $\varepsilon_- = 10$. The other parameters are the same as in Fig.~\ref{fig:Drude-dimer-dipole}.}
\end{figure}

Figure~\ref{fig:Drude-dimer-dipole-h-scan} shows how the response of the dimer depends on the distance $h$ above the substrate. A substrate dielectric function $\varepsilon_- = 10$ was assumed in order to emphasize the influence of the substrate on the response of the dimer. In Fig.~\ref{fig:Drude-dimer-dipole-h-scan}(a), where $h = 2a$, the spectral response is very close to that of the isolated dimer, shown in Fig.~\ref{fig:Drude-dimer-dipole}(a). We see clear interaction effects with the substrate when $h = 0.3a$ [Fig.~\ref{fig:Drude-dimer-dipole-h-scan}(b)], but it remains true that the spectrum is qualitatively similar to that of the free dimer. The dimer has to be close to the substrate for the interaction effects to modify the spectrum even for the large substrate dielectric function used in these calculations [Fig.~\ref{fig:Drude-dimer-dipole-h-scan}(c)]. 

\begin{figure}[thbp]
    \centering
    \subfigure[$\hbar\omega = 1.35$ eV.]{
        \includegraphics{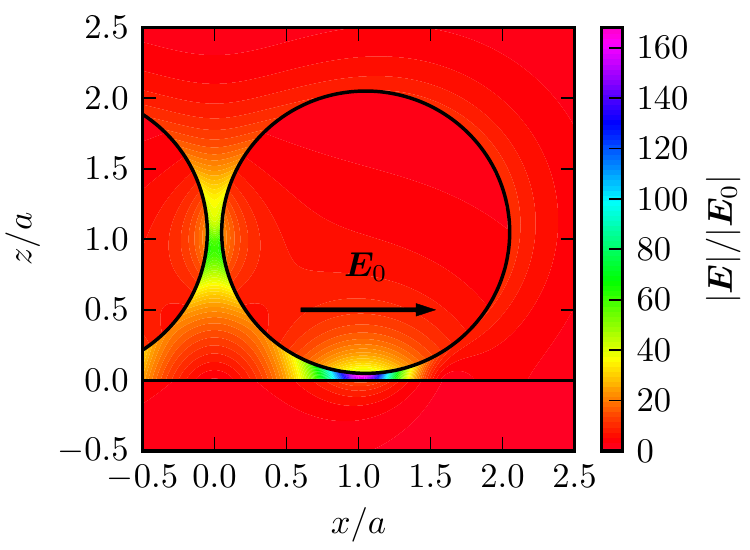}
    }
    \subfigure[$\hbar\omega = 1.45$ eV.]{
        \includegraphics{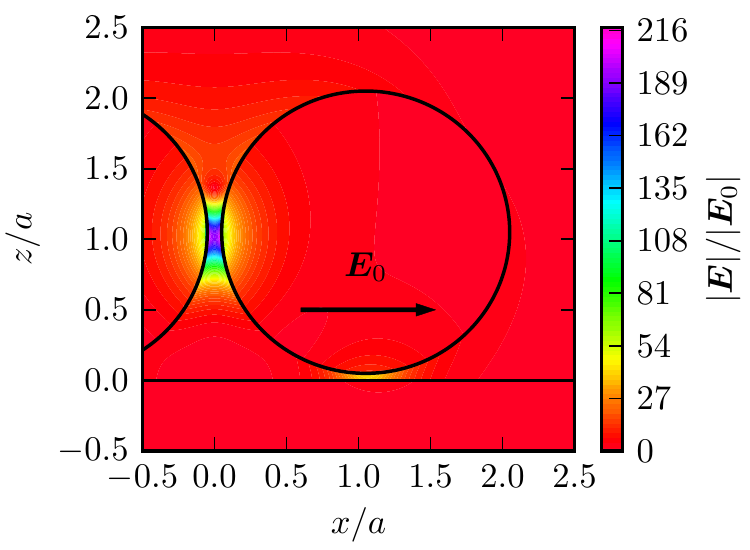}
    }
    \caption{\label{fig:Drude-field-enhancement}
    (Color online)
    The field enhancement $|\bm{E}|/|\bm{E}_0|$ in the $xz$ plane. At $\hbar\omega = 1.35$ eV the highest field enhancement is found between the sphere and the substrate; at $\hbar\omega = 1.45$ eV the highest field enhancement is found on the line connecting the two spheres. As shown in the figure, $\bm{E}_{0}\parallel\bm{\hat{x}}$. The system parameters are $\varepsilon_- = 10$, $h = 0.05 a$, $d = 0.1 a$, and $L = 50$. Of particular interest is the area between the two lowest-frequency peaks, where the location of maximum field enhancement ``flips'' between the two points indicated.}
\end{figure}

We pause for a moment to comment on issues of convergence. It is common to employ the dipole approximation to describe intersphere interactions, and also interactions of nanoparticles with substrates using the image method~\cite{Yamaguchi1974173}.
In Fig.~\ref{fig:Drude-dimer-h-scan-Ez} we present how the position of the lowest energy collective mode for the case $\bm{E}_{0}\parallel\bm{\hat{z}}$ depends on $L$, which determines the number of unknown $A_{lm}$ coefficients in Eq.~\eqref{eq:eq-system}. The frequently used dipole approximation corresponds to $L = 1$, and from Fig.~\ref{fig:Drude-dimer-h-scan-Ez} one observes that it is inaccurate even when the dimer is far above the substrate ($h = 2a$), and becomes gradually worse as $h$ is decreased. The cutoff $L$ must be on the order of $30$ to obtain converged results for the parameter ranges explored in this paper.
One may appreciate the reason for this from earlier work~\cite{Chu:2008gd}. When two spheres are quite close to each other, one encounters collective modes wherein the fields are concentrated in a small angular range near the points of closest contact. Similarly, when one or more nanospheres are very close to a dielectric substrate, one encounters collective modes localized around the ``south pole'' of the spheres---the points closest to the substrate.
One requires large values of the cutoff $L$ if one wishes to describe such modes accurately. Notice, by the way, that the mode frequency is significantly redshifted when the dimer comes very close to touching the substrate. 
In passing, we note that convergent results do not guarantee correctness of the calculated potentials. In order to do so, one has to explicitly make sure that the boundary conditions are satisfied to the required accuracy at all points on all interfaces~\cite{PhysRevB.61.772}.

In Fig.~\ref{fig:field-enhancement-Drude-gap}, we examine the nature of the enhanced fields in the Drude dimer at the two points indicated in the inset. Again we have assumed $d = 0.1a$ and $h = 0.05a$ in these calculations. For most of the spectrum, the largest field enhancement is found at point $2$, the hot-spot where the two spheres nearly touch. Notice, however, that we have very large field enhancements also between the south pole of the spheres and the dielectric substrate, in particular at $1.35$ eV [see Figs.~\ref{fig:field-enhancement-Drude-gap} and~\ref{fig:Drude-field-enhancement}(a)]. In our view, the region where the dielectric substrate is very close to the bottom of the sphere acts like an effective potential well which ``traps'' surface plasmons at the south pole. The surface plasmons sense the presence of the dielectric through the fields associated with them in the region outside the sphere.      

We can see from Fig.~\ref{fig:field-enhancement-Drude-gap} that point $1$ and point $2$ are ``hot'' simultaneously at roughly the same frequency. However, as one scans through a given resonance peak, near $\hbar\omega = 1.4$ eV in Fig.~\ref{fig:field-enhancement-Drude-gap}, the hot-spot moves from point $1$ to point $2$ and conversely, depending on the precise value of the frequency. Thus we have another example of the phenomenon of the ``moving hot-spots'' discussed in a recent publication~\cite{Letnes:2010uq}. We illustrate this behavior in Fig.~\ref{fig:Drude-field-enhancement}, where we plot $|\bm{E}| / |\bm{E}_0|$ in a contour map\footnote{Note that we do not plot the square of $|\bm{E}| / |\bm{E}_0|$, as the enhancement then becomes very large, in which case it is difficult to synthesize a nicely color coded figure.}. A small energy shift of $0.1$ eV is enough to change the shape of the field enhancement considerably, and move the hot spot from point 1 to point 2.

% subsection The Drude Monomer and Dimer (end)

\subsection{Ag Monomers and Dimers} % (fold)
\label{sub:ag-monomers-and-dimers}

\begin{figure}[bthp]
    \centering
    \includegraphics{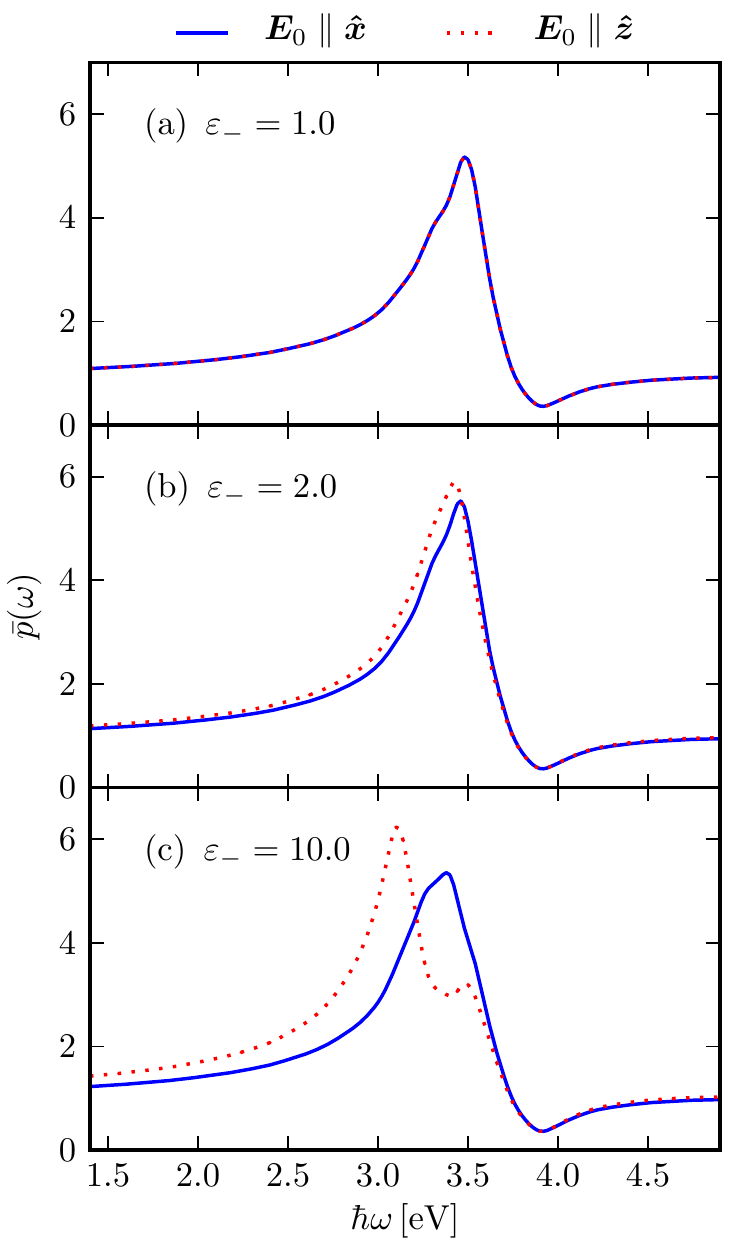}
    \caption{\label{fig:Ag-monomer-dipole}
    (Color online)
    The dimensionless dipole moment for an Ag sphere placed a distance $h = 0.05a$ above a dielectric substrate of (a) $\varepsilon_- = 1$, (b) $\varepsilon_- = 2$, and (c) $\varepsilon_- = 10$. The equation system was truncated at $L = 50$.
    }
\end{figure}

The Drude model discussed in Sec.~\ref{sub:drude-monomer-dimer} is useful to examine, since one may model metals in which the plasmons are damped very lightly. Thus one can explore detailed structure in the response of the model system. In practice, however, interest resides in realistic metals that display plasmonic response in the visible. In this respect, silver (Ag) and gold (Au) are the two metals most studied experimentally. While Au is indeed ``plasmon active'', the plasmons in this material are in fact rather heavily damped. Ag is a much better material in principle, even though in experiments oxide can form on its surface.

\begin{figure}[bthp]
    \centering
    \includegraphics{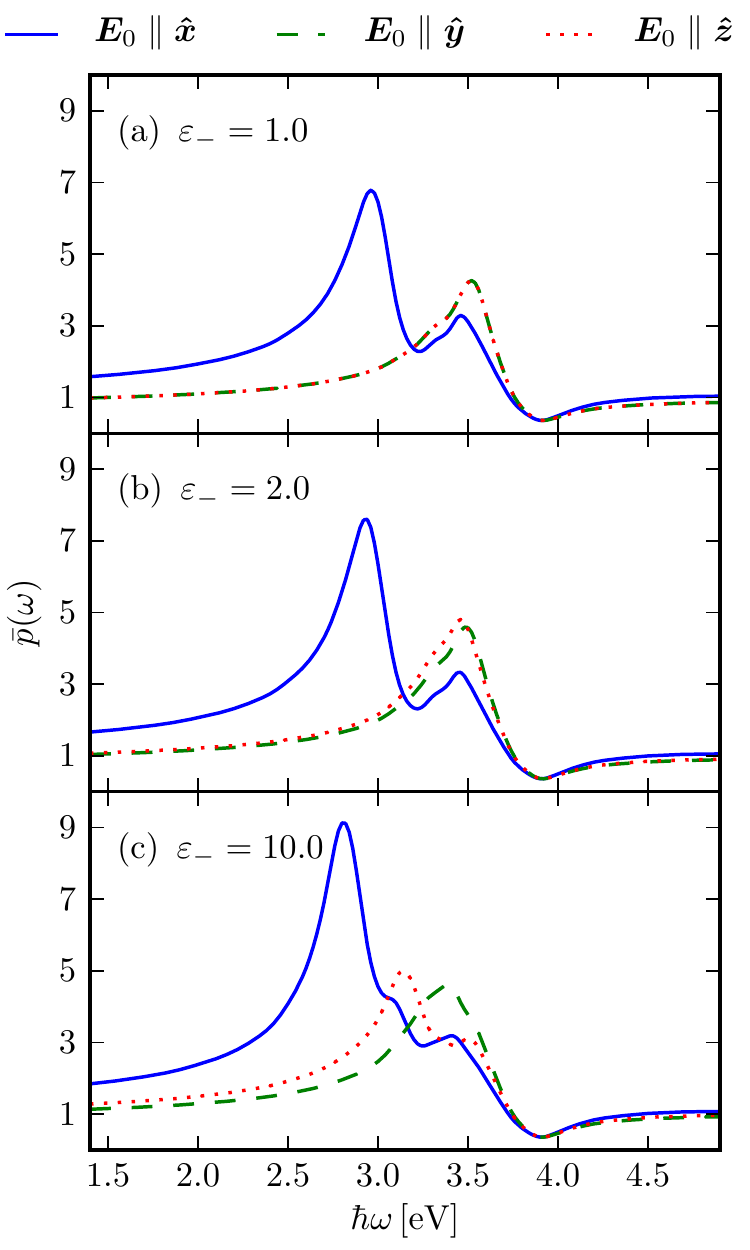}
    \caption{\label{fig:Ag-dimer-dipole}
    (Color online)
    The response for an Ag dimer close to a substrate, as described by the dimensionless dipole moment of one of the spheres. As in previous cases, $d = 0.1a$, $h = 0.05a$, and $L = 50$ were used.
    }
\end{figure}
This section is devoted to studies of the plasmon resonance properties of Ag monomers and dimers. Figure~\ref{fig:Ag-monomer-dipole} shows calculations of the reduced dipole moment for a single Ag nanosphere placed a distance $h = 0.05a$ over a dielectric substrate. For a free-standing Ag sphere in vacuum, the Mie resonance at $\hbar\omega = 3.5$ eV is readily observed [Fig.~\ref{fig:Ag-monomer-dipole}(a)]. The response of the sphere is modest for $\varepsilon_- = 2$ [Fig.~\ref{fig:Ag-monomer-dipole}(b)]. However, when $\varepsilon_- = 10$ [Fig.~\ref{fig:Ag-monomer-dipole}(c)], we see a substantial splitting of the main resonance and activation of higher frequency modes occur.

In Fig.~\ref{fig:Ag-dimer-dipole} we show the response of an Ag dimer with its axis parallel to the substrate. As previously, we have assumed $h = 0.05a$ and $d = 0.1a$. The strong interaction between the two spheres of a free-standing dimer in vacuum can be seen from Fig.~\ref{fig:Ag-dimer-dipole}(a) by noting the pronounced difference in response to an applied field parallel ($\bm{E}_{0}\parallel \bm{\hat{x}}$) or perpendicular ($\bm{E}_{0}\parallel \bm{\hat{y}}$ or $\bm{E}_{0}\parallel \bm{\hat{z}}$) to the dimer axis. At least from the perspective of the dipole moment of each sphere, the influence of the substrate is not significant for $\varepsilon_- = 2$ [Fig.~\ref{fig:Ag-dimer-dipole}(b)], but we see substantial effects for the larger dielectric function, $\varepsilon_- = 10$ [Fig.~\ref{fig:Ag-dimer-dipole}(c)].

\begin{figure}[tbhp]
    \centering
    \includegraphics{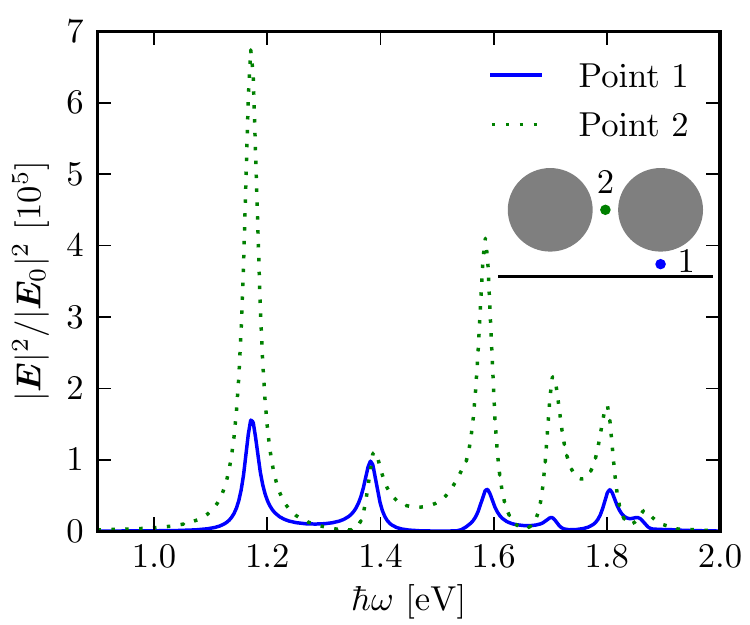}
    \caption{\label{fig:field-enhancement-Ag-gap}
    (Color online)
    Intensity enhancement, $|\bm{E}|^2 / |\bm{E}_0|^2$, as a function of frequency of the applied field at point 1 and point 2, for the Ag dimer on a substrate of dielectric function $\varepsilon_- = 10$. The remaining parameters are the same as in Fig.~\ref{fig:Ag-dimer-dipole}.
    }
\end{figure}

While the dipole moment of the Ag spheres shows substrate effects to be weaker than those of the corresponding Drude monomer and dimer, the field enhancement effects are still substantial. When the spheres are either close to each other and/or close to the substrate, the resonances are highly localized in space and form so-called ``hot-spots''. This is illustrated by Fig.~\ref{fig:field-enhancement-Ag-gap}, which shows the enhancement in the electric field intensity ($|\bm{E}|^2 / |\bm{E}_0|^2$) for an Ag dimer. Hence, one can have local regions where the fields are strongly enhanced while their effect on the total dipole moment of the sphere is more modest.

Regarding the field enhancement in the Ag dimer, depicted in Fig.~\ref{fig:field-enhancement-Ag-gap}, we see considerable enhancement between the sphere and the substrate. This enhancement is caused by the proximity of the sphere to the dielectric substrate that creates a potential well where surface plasmons can be trapped near the south pole of the sphere. On resonance, the enhancement in the square of the field is close to $5 \times 10^3$ [Fig.~\ref{fig:field-enhancement-Ag-gap}]. If one has SERS in mind, where the cross section is enhanced by roughly the fourth power of the field, then the Raman cross section would in this case be enhanced by $25 \times 10^6$.  Thus, the influence of the dielectric substrate on the enhanced fields realized for the dimer is very substantial. Although we observed full reversal of hot-spot positions for a dimer made from Drude metal, it appears that the larger attenuation of silver ($\Im [\varepsilon_j(\omega)]$) prohibits this phenomena in the Ag dimer. Hence, the dominant hot-spot is for all frequencies of the incident light located at point 2, in the gap between the two spheres.

%\begin{figure}[tbhp]
    %\centering
    %\subfigure[$\hbar\omega = 3.04$ eV.]{
        %\includegraphics{figures/Eabs-contour-Ag-dimer-on-10.0/3_04eV}
    %}
    %\subfigure[$\hbar\omega = 3.26$ eV.]{
        %\includegraphics{figures/Eabs-contour-Ag-dimer-on-10.0/3_26eV}
    %}
    %\caption{\label{fig:Ag-field-enhancement}
        %The field enhancement $|\bm{E}|/|\bm{E}_0|$ in the $xz$ plane for an Ag dimer on a dielectric substrate with $\varepsilon_- = 10$. While the field enhancement between the particle and the substrate is not as strong as for the Drude metal, we see that the particle-substrate hot-spot (located directly below the Ag particle)  varies greatly with a small change in frequency of the incident light. Distance parameters are $d = 0.05 a$ and $h = 0.1 a$. TODO: should we just remove this figure? it is not as interesting as the Drude case anyway, maybe fig. 10 is enough on this matter.}
%\end{figure}

% subsection Ag Monomers and Dimers (end)

\begin{figure}[bthp]
    \centering
        \includegraphics{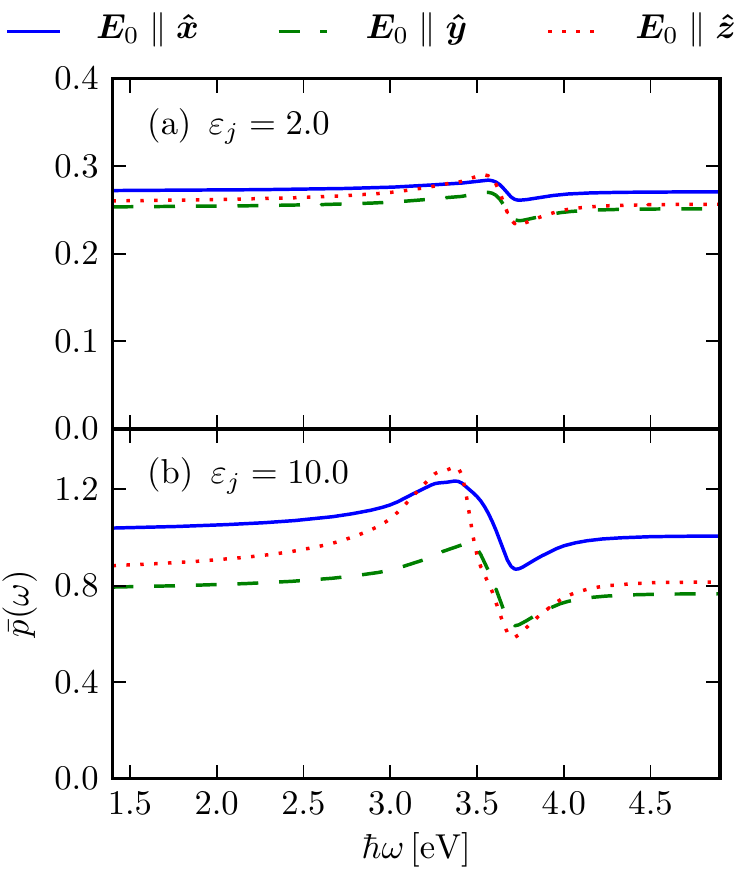}
    \caption{\label{fig:dielectric-dimer-on-Ag-dipole}
    (Color online)
    The dimensionless dipole moment of a member of a dielectric dimer placed in close proximity to an Ag surface. The dielectric functions for the two spheres forming the dimer are both (a) $\varepsilon_j = 2$ and (b) $\varepsilon_j = 10$. In both cases we have $d = 0.1a$, $h = 0.05a$, and $L = 50$.}
\end{figure}

\subsection{Dielectric Dimer on an Ag Substrate} % (fold)
\label{sub:dielectric-dimer-ag-substrate}

In the literature, primary attention is directed toward nano-scale objects fabricated from plasmon active metals. We find that dielectric particles create localized evanescent fields that stimulate the formation of localized plasmons in a nearby metallic substrate. These surface plasmons are not excited on a flat metallic surface. In this section we consider a dielectric dimer with a frequency independent, real and positive dielectric function $\varepsilon_j$ placed close to an Ag surface.

Figure~\ref{fig:dielectric-dimer-on-Ag-dipole} depicts the dimensionless dipole moment of the dielectric dimers placed very close to an Ag substrate. As before, the separation between the two spheres is $d = 0.1a$ and the height above the substrate is $h = 0.05 a$. Dipole activity in the dielectric sphere is observed in the frequency range near the surface plasmon resonance of the Ag surface. Since the dielectric function of the sphere ($\varepsilon_j$) is frequency independent, this plasmonic activity has its origin in the Ag substrate. As expected, the effect is enhanced when the dielectric function of the spheres, $\varepsilon_j$, is increased [Fig.~\ref{fig:dielectric-dimer-on-Ag-dipole}].

\begin{figure}
    \centering
    \includegraphics{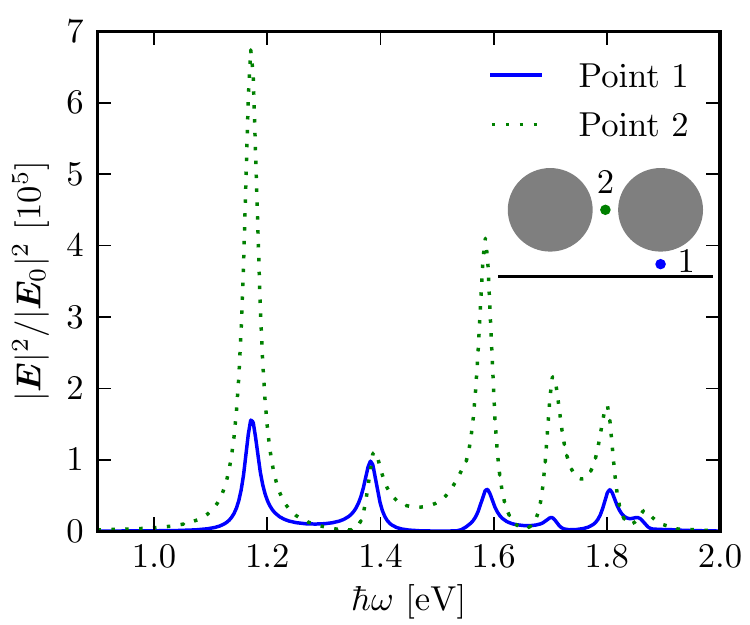}
    \caption{\label{fig:dielectric-dimer-field-enhancement}
    (Color online)
    Intensity enhancements at two selected points for a dielectric dimer placed in the near vicinity of an Ag surface. The dielectric function of the spheres is $\varepsilon = 10$, and other parameters are the same as in Fig.~\ref{fig:dielectric-dimer-on-Ag-dipole}.}
\end{figure}

\begin{figure}[thbp]
    \centering
    \subfigure[$\hbar\omega = 3.02$ eV.]{
        \includegraphics{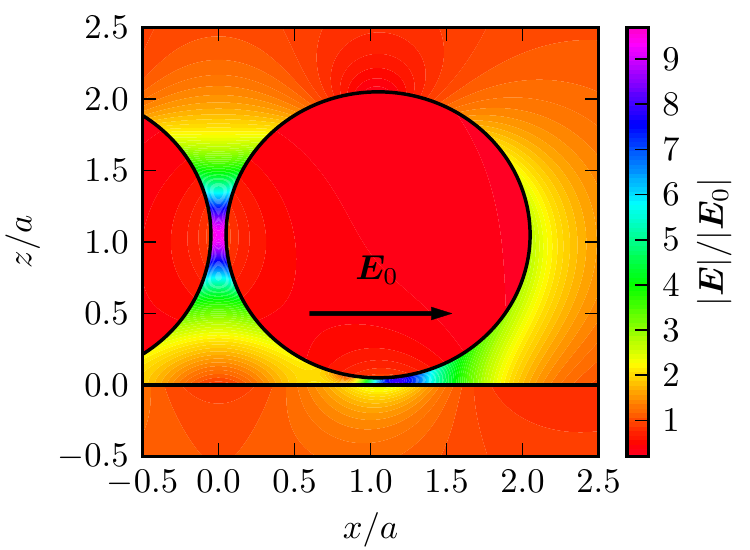}
    }
    \subfigure[$\hbar\omega = 3.39$ eV.]{
        \includegraphics{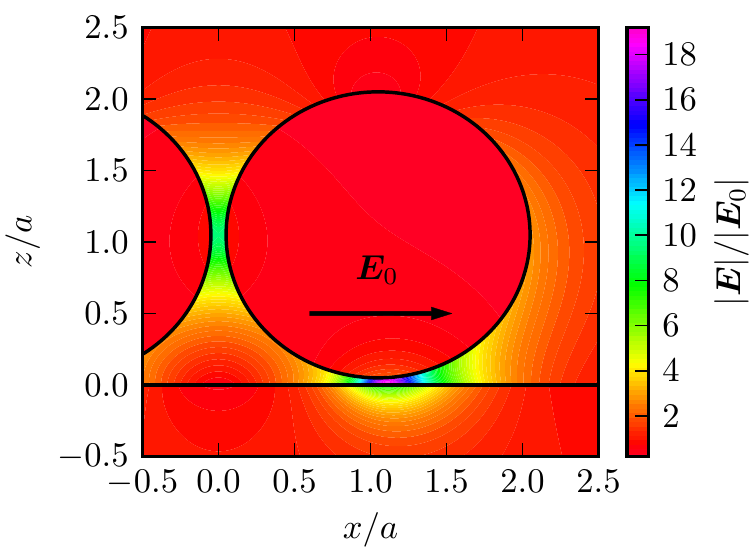}
    }
    \caption{\label{fig:dielectric-field-enhancement}
    (Color online)
    The field enhancement $|\bm{E}|/|\bm{E}_0|$ in the $xz$ plane for a dielectric dimer above an Ag substrate. At $\hbar\omega = 3.02$ eV the highest field enhancement is found between the spheres; at $\hbar\omega = 3.39$ eV the highest field enhancement is found on the line connecting the two spheres. As shown in the figure, $\bm{E}_{0}\parallel\bm{\hat{x}}$. The parameters are $\varepsilon_j = 10$, $h = 0.05 a$, $d = 0.1 a$, and $L = 50$.}
\end{figure}

Figure~\ref{fig:dielectric-dimer-field-enhancement} illustrates the frequency dependence of the intensity enhancement at a position between the dielectric spheres and the substrate (point 1) and between the spheres (point 2). Between the spheres (point $2$) the plasmonic response of the substrate plays only a minor role in the intensity enhancement, whereas just below the south pole of the spheres, the plasmonic activity plays a more important role, producing higher intensity enhancement for a narrow part of the spectrum. The proximity of the dielectric spheres to the substrate converts the incoming plane wave to an evanescent wave which excites surface plasmons in the Ag substrate. The consequence is that substantial intensity enhancements appear below the nanospheres near the surface plasmon frequency of the Ag surface. The physics is quite similar to the formation of a hot-spot between a metal sphere and a dielectric substrate.

The structure of the field enhancement near the plasmon resonance is shown in detail in Fig~\ref{fig:dielectric-field-enhancement}. This figure is qualitatively similar to as Fig.~\ref{fig:Drude-field-enhancement}. Again, we are faced with the moving hot-spot phenomenon, since the hot-spot moves from between the spheres to below the spheres as the frequency of the incident light changes.

% subsection Dielectric Dimer on an Ag Substrate (end)

% section Results and discussion (end)

\section{Concluding remarks} % (fold)
\label{sec:Concluding-remarks}

We have formulated the theory of the interaction of non-periodic nanosphere arrays with a substrate, and for the case of monomers and dimers we have provided numerical studies of electric dipole moments induced by a uniform driving field, and also field enhancement generated by excitation of plasmon resonances. While the focus usually is placed on the interaction of metallic nanoscale objects with metallic substrates or surroundings, our emphasis has been on the interaction between metallic and dielectric materials: metallic nanoparticles on dielectric substrates, or dielectric nanoparticles on metallic substrates. Such systems are, in our view, better suited for experimental examination.

For both these configurations we find ``hot-spots'' (i.e. local regions of high intensity) that are localized between the south pole of the nanosphere and the substrate. The physical origin of this behavior is an effective potential well created by the dielectric that traps and localizes plasmons in the nearby metallic component. Consider, for example, a semi-infinite slab of a model metal described by the Drude model, Eq.~\eqref{eq:drudemodel}. Let the metal lie in the half space $z < 0$, and let the half space $z > 0$ be vacuum. The surface supports surface plasmons, and in electrostatic theory these have frequency $\omega_P / \sqrt{2}$ independent of wave vector. Suppose we instead fill the upper half space $z > 0$ with a dielectric material whose dielectric constant is $\varepsilon_+ > 1$. This lowers the frequency of surface plasmons on the metal surface to $\omega_P (1 + \varepsilon_+)^{-1/2}$.
If we then imagine that the dielectric covers only a finite area on the metallic surface, clearly an attractive potential well is formed which can trap surface plasmons bound to the region where the dielectric is found. The frequency of these modes lies below the frequency band associated with those on the metal/vacuum interface. In our studies, we have a rather different geometry. For instance, in one configuration we explored a dielectric sphere that is placed just a bit above the metallic substrate.
The surface plasmons on the metal surface ``sense'' the presence of the dielectric through their evanescent field that extends above the metal surface. We then find plasmon modes localized in the near vicinity of the south pole of the sphere. In the case of dimers, of interest is the ``moving hot-spot'' phenomenon illustrated in Fig.~\ref{fig:Drude-field-enhancement}. Small changes in excitation frequency results in a hot-spot that moves from one point in the structure to another. An earlier discussion provided an example of this behavior in a rather different structure~\cite{Letnes:2010uq}. 

The hot-spots localized between metallic spheres and dielectric substrates, and also between dielectric spheres and metallic substrates, suggests that strongly enhanced non-linear optical studies may be carried out on diverse systems, not just those where all constituents are plasmon active metals. It would be of interest to explore field enhancements not just for spheres placed near flat substrates, but for other nanoscale objects of diverse shape as well. It should be possible to engineer structures in which large field enhancements are realized that can be exploited to study, for instance, adsorbates on insulating surfaces. 

% section Concluding remarks (end)

\begin{acknowledgments}

We used the freely available software archive SHTOOLS authored by Mark Wieczorek (available at \url{http://www.ipgp.fr/~wieczor/SHTOOLS/SHTOOLS.html}) to evaulate the associated Legendre functions.

P.A.L. and I.S. would like to thank the Department of Physics and Astronomy, University of California at Irvine for kind hospitality. P.A.L. would like to thank Dr. Ping Chu and Tor Nordam for interesting discussions. He would also like to thank ``NTH-fondet'' for its support of his research.  The research of D.L.M. has been supported by the U. S. National Science Foundation, through grant No. CHE-0533162.  I.S. acknowledges the support from the Research Council of Norway under the program Sm\aa{}forsk and an NTNU Mobility Fellowship.

\end{acknowledgments}

\appendix
\section{Construction of the equation system} % (fold)
\label{sec:Construction the equation system}

The expression for the electrostatic potential, $\psi(\bm{r})$, is shown in Eqs.~\eqref{eq:plus-expansion} and~\eqref{eq:psi_j}. This expansion is of no use before one has determined the expansion coefficients $A_{lm}^{(j)}$ and $B_{lm}^{(j)}$. To do so, we combine the series expansion of $\psi$ and require the fulfillment of the boundary conditions at the surface of the spheres, i.e. the continuity of $\psi$ and $\varepsilon \partial_n \psi$ over any interface (where $\partial_n = \bm{\hat{n}} \cdot \bm{\nabla}$ denotes the normal derivative)~\cite{Jackson:1962aa}. Note that the boundary conditions at the interface at $z = 0$ are already fulfilled through Eq.~\eqref{eq:image-bc}. For instance we may consider continuity of the electrostatic potential at the surface of sphere $j$. This gives the condition 
\begin{align}
\label{eq:boundary-condition}
    \begin{aligned}
        \lim_{r_j \rightarrow a_j^-} &\psi_j(\bm{r}_j)
        =
        \lim_{r_j \rightarrow a_j^+}
        \Bigg\{
            -\bm{r} \cdot \bm{E}_0 \\
            &+ \sum_{i=1}^{N} \psi_i(\bm{r}_i)
                + \sum_{i=1}^{N} \psi_{\bar{i}}(\bm{r}_i)
        \Bigg\},
    \end{aligned}
\end{align}
where the notation $a_j^\pm$ means $a_j \pm \eta$, where $\eta$ is infinetesimally small and positive.
For this condition to be useful, and similarly for the equation following from the continuity of the normal components of the displacement field $\bm{D} = -\varepsilon \varepsilon_0 \bm{\nabla} \psi$, we need to express the potentials $\psi_{i}$ and $\psi_{\bar{i}}$ for $i \neq j$ in terms of the coordinate system $\mathcal{S}_j$ centered on sphere $j$. One can do this by using an identity employed by Bedeaux and Vlieger~\cite{bedeaux_book}. This reads
\begin{align}
\label{eq:ylm-translation}
    \begin{aligned}
        r_i^{-{l_i}-1} & Y_{l_i}^{m_i}(\theta_i, \phi_i) =
            \sum_{l_j = 0}^\infty \sum_{m_j = -l_j}^{l_j}
                H(l_j, m_j, | l_i, m_i) \\
                &\times
                \frac
                    {Y_{l_i + l_j}^{m_i - m_j} (\theta_{ij}, \phi_{ij})}
                    {R_{ij}^{l_j + l_i + 1}}
                r_j^{l_j} Y_{l_j}^{m_j} (\theta_j, \phi_j),
    \end{aligned}
\end{align}
where $\bm{R}_{ij}$ is the vector between the center of the sphere $i$ and $j$, and $\theta_{ij}$ and $\phi_{ij}$ are the polar and azimuthal angles, respectively, which describe the direction of $\bm{R}_{ij}$. In writing Eq.~\eqref{eq:ylm-translation}, we have used
\begin{align}
\label{eq:H-def}
    \begin{aligned}
        H (l_j, m_j, | & l_i, m_i) =
        \sqrt{4\pi} (-1)^{l_i + m_j} \\
        & \times \left[
            \frac
                {2 l_i + 1}
                {(2 l_j + 1)(2l + 1)}
        \right]^{1/2} \\
        & \times
        \left[
            \binom{l + m}{l_i + m_i}
            \binom{l - m}{l_j + m_j}
        \right]^{1/2},
    \end{aligned}
\end{align}
where $l = l_j + l_i$ and $m = m_i - m_j$. Moreover, the notation $\binom{a}{b}$ denotes the binomial coefficient. The expansion described by Eqs.~\eqref{eq:ylm-translation} and ~\eqref{eq:H-def} can also be applied to the image multipoles located in the substrate ($z < 0$). 

We now have the electrostatic potential on each side of the surface of sphere $j$ expressed in terms of the coordinates of system $\mathcal{S}_j$. One may generate a system of equations for the unknown amplitudes by equating the coefficients of $Y_{l_j}^{m_j}(\theta_j, \phi_j)$. When Eq.~\eqref{eq:boundary-condition} is combined with the condition that the normal components of the electric displacement field $\bm{D}$ should be continuous across the surface of sphere $j$, it is possible to eliminate the coefficients $B_{lm}^{(j)}$ and to generate a linear system of equations that involves only $A_{lm}^{(j)}$. When this is done, the following linear system of equations results:
\begin{widetext}
\begin{align}
    \label{eq:eq-system}
    \begin{aligned}
    % RHS first
    - b_{1m_j} & \delta_{1, l_j}
    =
    % LHS below
    A^{(j)}_{l_j m_j}
    \frac {l_j \varepsilon_j + \varepsilon_+ (l_j + 1)}
        {l_j (\varepsilon_j - \varepsilon_+)}
    a_j^{-2 l_j-1}
    -\sum_{l_j , m_j}
        A_{l_i m_i}^{(j)}
        H(l_j, m_j | l_i, m_i)
            \frac
                {Y_{l_i + l_j}^{m_i - m_j} (\theta_{ij}, \phi_{ij})}
                {R_{ij}^{l_j + l_i + 1}} \\
    %
    %& + \sum_{l_{\bar{j}} \geq |m_j|}
        %A_{l_{\bar{j}} m_j}^{(j)}
        %(-1)^{l_j + m_j} \beta
        %H(l_j, m_j | l_{\bar{j}}, m_j)
        %\sqrt{\frac{2 (l_j + l_{\bar{j}}) + 1}{4\pi}}
        %\frac
            %{1}
            %{R_{\bar{j}j}^{l_j + l_{\bar{j}} + 1}} \\
    %
    &+ \sum_{i\neq j} \sum_{l_i , m_i}
        A_{l_i m_i}^{(i)}
        H(l_j, m_j | l_i, m_i)
        \Bigg[
            \frac {Y_{l_i + l_j}^{m_i - m_j} (\theta_{ij}, \phi_{ij})}
                {R_{ij}^{l_j + l_i + 1}}
            + (-1)^{l_i + m_i}
        \beta
        \frac
            {Y_{l_i + l_j}^{m_i - m_j}
                (\theta_{\bar{i}j}, \phi_{\bar{i}j})}
            {R_{\bar{i}j}^{l_j + l_i + 1}}
        \Bigg],
    \end{aligned}
\end{align}
\end{widetext}
where $\beta = (\varepsilon_+ - \varepsilon_-) / (\varepsilon_+ + \varepsilon_-)$ and $l_j = 1, 2, 3, \ldots, L$ and $m_j = 0, \pm 1, \pm 2, \ldots, \pm l_j$. The coefficients $b_{lm}$ are the expansion coefficients of the applied field $\bm{E}_0$ in terms of the spherical harmonics. The non-zero $b_{lm}$ coefficients (in the case of uniform $\bm{E}_0$) are given by~\cite{bedeaux_book,PhysRevB.61.772}
\begin{subequations}
    \begin{align}
        b_{10} &= -E_0 \sqrt{\frac{4\pi}{3}} \cos\theta_0, \\
        b_{1\pm 1} &= \pm E_0 \sqrt{\frac{2\pi}{3}}
            \sin\theta_0 \mathrm{e}^{\mp \mathrm{i} \phi_0},
    \end{align}
\end{subequations}
where $\theta_0$ is the angle between the external field and the positive $z$ axis and $\phi_0$ is the azimuthal angle which describes the angle between the projection of the external field onto the $xy$ plane and the positive $x$ axis. As the Laplace equation is linear, we only need to solve for 3 different directions of $\bm{E}_0$ ($\bm{E}_0$ parallel to $\bm{\hat{x}}, \,\, \bm{\hat{y}}, \text{ and } \bm{\hat{z}}$). The response to an applied field pointing in any other direction can be constructed through superposition of these 3 cases.

Equation~\eqref{eq:eq-system} gives us $N (L + 1)^2 - 1$ linear equations in the expansion coefficients $A_{lm}^{(j)}$, and we have $N (L + 1)^2$ unknowns. The final equation results from the continuity of the normal component of $\bm{D}$ at the spherical interfaces. Taking the normal (i.e. radial) derivative of the $B_{00}$ term, we see that this term vanishes ($\partial_{r_j} B_{00} Y_0^0(\theta_j, \phi_j) = 0$). This means that $A_{00} = 0$, related to the fact that the nanoparticles are assumed to carry no charge. Hence, the equation system is closed, and we can expect to find a unique solution.

Finally, the resonances for an isolated sphere in a homogeneous background of dielectric function $\varepsilon_+$ can be obtained from Eq.~\eqref{eq:eq-system}. By neglecting all contributions from other particles and image multipoles, i.e. to keep only the first term on the right hand side of Eq.~\eqref{eq:eq-system}, one is essentially left with the isolated sphere case. Under this assumption, the resulting equation can readily be solved to give
\begin{align*}
    A_{lm} \propto \frac{1}{l \varepsilon + \varepsilon_+ (l + 1)}.
\end{align*} 
where $\varepsilon$ is the dielectric function of the sphere.
Hence, the resonance positions are determined by the zeroes of the real part of the denominator of $A_{lm}$:
\begin{align*}
    \Re[l \varepsilon + \varepsilon_+ (l + 1)] = 0.
\end{align*}
If we assume for $\varepsilon$ the Drude model with $\gamma = 0$ ($\varepsilon = 1 - \omega_P^2 / \omega^2$), which in our case is a good approximation, we get the following resonance frequencies for the isolated sphere:
\begin{align*}
    \omega_l = \omega_P \sqrt{\frac{l}{2l + 1}}.
\end{align*}
For systems containing more than a single, isolated sphere, such as the ones discussed in this paper, these resonance frequencies are typically modified due to particle-particle or particle-substrate interactions.

% section Construction the equation system (end)

\bibliography{references}

\end{document}